\documentclass[aps,prd,amsmath,amssymb,
preprintnumbers,nofootinbib]{revtex4-1}

\usepackage{graphicx}
\usepackage{dcolumn}
\usepackage{bm}
\usepackage[bookmarks, pagebackref=false]{hyperref}
\usepackage[usenames,dvipsnames]{xcolor}
\definecolor{rossoCP3}{cmyk}{0,.88,.77,.40}
\definecolor{blaa}{rgb}{0.2,0.2,0.6}
\hypersetup{colorlinks, 
	bookmarksopen, 
	bookmarksnumbered,
	citecolor=blaa, 		
	linkcolor=rossoCP3,	
	urlcolor=rossoCP3,			
}

\usepackage{amsmath}
\usepackage{slashed}
\usepackage{tikz}
\usepackage{graphicx}
\usepackage{pgfplots}
\usepackage{subfigure}
\usepackage{color}
\usepackage{rotating}
\usepackage{gensymb}
\usepackage{longtable}
\usepackage{capt-of}
\usepackage{simplewick}
\usepackage{enumerate}
\usetikzlibrary{quotes,angles,snakes}
\usetikzlibrary{trees}
\usetikzlibrary{decorations.pathmorphing}
\usetikzlibrary{decorations.markings}
\usetikzlibrary{arrows,shapes,positioning}
\usetikzlibrary{decorations.markings}
\tikzstyle arrowstyle=[scale=1]
\tikzstyle directed=[postaction={decorate,decoration={markings,
    mark=at position .65 with {\arrow[arrowstyle]{stealth}}}}]
\tikzstyle reverse directed=[postaction={decorate,decoration={markings,
    mark=at position .65 with {\arrowreversed[arrowstyle]{stealth};}}}]

\newcommand{\RNum}[1]{\uppercase\expandafter{\romannumeral #1\relax}}

\newcommand{\beq}{\begin{equation}}
\newcommand{\eeq}{\end{equation}}
\newcommand{\bea}{\begin{eqnarray}}
\newcommand{\eea}{\end{eqnarray}}

\usepackage{xcolor,colortbl}
\definecolor{Blue}{RGB}{140,165,195}
\definecolor{Purple}{RGB}{255,145,145}
\definecolor{bluc}{cmyk}{1,1,0,0.1}
\definecolor{rossoCP3}{cmyk}{0,.88,.77,.40}
\definecolor{rosso}{cmyk}{0,1,1,0.4}
\definecolor{rossos}{cmyk}{0,1,1,0.55}
\definecolor{rossoc}{cmyk}{0,1,1,0.2}
\definecolor{verdes}{cmyk}{0.92,0,0.59,0.4}

\newcommand{\MSRG}{{\rm MSRG\!}}
\newcommand{\CWRG}{{\rm CWRG\!}}
\newcommand{\MS}{{\rm MS}}
\newcommand{\CW}{{\rm CW}}

\usepackage{color}

\begin{document}

\newcommand{\blue}[1]{{\color{blue}#1}}
\newcommand{\red}[1]{{\color{red}#1}}
 
\title{ \LARGE  \color{rossoCP3}  Transformation 
of scalar couplings
between Coleman-Weinberg and 
MS
schemes 
}
\author{F.A. Chishtie$^{\color{rossoCP3}{\triangle}}$}
\author{Zhuo-Ran Huang$^{\color{rossoCP3}{\clubsuit}}$}
\author{M. Reimer$^{\color{rossoCP3}{\diamondsuit}}$}
\author{T.G. Steele$^{\color{rossoCP3}{\diamondsuit}}$}
\author{Zhi-Wei Wang$^{\color{rossoCP3}{\heartsuit}}$}
\affiliation{$^{\color{rossoCP3}{\diamondsuit}}${\mbox {Department of Physics and Engineering physics, University of Saskatchewan, Saskatoon, SK S7N 5E2, Canada}}\\
$^{\color{rossoCP3}{\heartsuit}}$ {\mbox {$\rm{CP}^3$-Origins, University of Southern Denmark, Campusvej 55}}
5230 Odense M, Denmark \\
$^{\color{rossoCP3}{\triangle}}$ {\mbox{Department of Physics and Astronomy, The University of Western Ontario, London, ON N6A 5B7, Canada}} \\
$^{\color{rossoCP3}{\clubsuit}}$ {\mbox{Universit\'e Paris-Saclay, CNRS/IN2P3, IJCLab, 91405 Orsay, France 
}} \\
}

\begin{abstract} 
The Coleman-Weinberg (CW) renormalization scheme for  renormalization-group improvement of the effective potential is particularly valuable for CW symmetry-breaking mechanisms (including the challenging case of models with multiple scalar fields). CW mechanism is typically studied using models with classical scale invariance which not only provide a possibility for an alternative symmetry breaking mechanism but also partially address the gauge hierarchies through dimensional transmutation.
 As outlined in our discussion section, when the couplings are not large, models with CW symmetry-breaking mechanisms have also been shown to naturally provide  the strong first-order phase transition necessary for stochastic gravitational wave signals.  A full understanding of the CW-MS scheme transformation of couplings thus  becomes important in the era of gravitational wave detection and precision  coupling measurements.
 A generalized Coleman-Weinberg (GCW) renormalization scheme is formulated  and methods for 
transforming scalar self-couplings between the GCW and MS (minimal-subtraction)  renormalization schemes are developed. 
Scalar $\lambda\Phi^4$ theory with global $O(4)$ symmetry is
explicitly studied up to six-loop order  to explore the magnitude  of this scheme transformation effect on the couplings. The dynamical rescaling of renormalization scales between the GCW and MS schemes 
can  lead to significant (order of 10\%) differences in the coupling at any order,  and consequently GCW-MS scheme transformation effects must be considered within precision  determinations of scalar couplings  
in extensions of the Standard Model. 
{\footnotesize  \it Preprint: CP$^3$-Origins-2020-03 DNRF90}
\end{abstract}
\maketitle

\section{Introduction}
The Coleman-Weinberg (CW) symmetry breaking mechanism \cite{Coleman:1973jx} has been an important research area from both theoretical and phenomenological perspectives. On the theory side, the  CW mechanism is the only known alternative symmetry breaking mechanism beyond conventional\footnote{Conventional in the sense of spontaneous symmetry breaking as Higgs mechanism.} symmetry breaking in the perturbative regime.\footnote{In the non-perturbative regime, there is also a strong dynamical symmetry breaking mechanism.} Also, studying the CW mechanism typically requires models with classical scale invariance, where similar to QCD, a hierarchy of the energy scales can be dynamically generated through dimensional transmutation, which partially addresses the well-known gauge hierarchy problem \cite{Weinberg:1978ym}.

From the phenomenological perspective, the era of gravitational wave detection and precision coupling measurements represents promising opportunities for observing new physics beyond the Standard model (SM).   Detection of stochastic gravitational wave signals typically require a strong first-order phase transition \cite{Grojean:2006bp} (see also \cite{Caprini:2019egz}). Contrary to the cases where the models can be engineered \footnote{The typical way of engineering the model is to increase the coupling of $\phi^3$ term by hand, where $\phi$ denotes a general scalar field.} to provide a strong first-order phase transition,
 models with Coleman-Weinberg (CW) symmetry breaking  can naturally\footnote{Natural is in the sense of without tuning the parameters by hand.} lead to this necessary strong first-order phase transition if the couplings are not too large\cite{Sher:1988mj,Sannino:2015wka,Huang:2020bbe}.  Thus models with CW symmetry-breaking  has been an interesting and meaningful  element in the study of strong first order phase transition and its associated stochastic gravitational waves.

On the other hand, Higgs cubic and quartic coupling measurements are extremely important in exploring the underlying mechanism of electroweak symmetry breaking (see e.g., typical Coleman-Weinberg type cases~\cite{Hill:2014mqa,Steele:2012av,Meissner:2006zh,Elias:2003zm,Dermisek:2013pta,Steele:2013fka,Wang:2015cda,Kanemura:2016tan} and implications of various forms of the effective potential~\cite{Agrawal:2019bpm}) and the nature of the electroweak phase transition ~\cite{Noble:2007kk}. The targeted sensitivity in  future collider experiments 
should have sufficient accuracy  to distinguish the conventional SM with most of the beyond-SM new physics
(see e.g., Refs.~\cite{CEPC-SPPCStudyGroup:2015csa,Liu:2018peg}). 
Thus it is important to understand how 
couplings in CW symmetry-breaking models  are related to  MS (minimal subtraction) scheme couplings and  observables, and to assess the magnitude of these scheme-transformation effects.

In the CW symmetry breaking mechanism, the associated 
CW renormalization scheme~\cite{Coleman:1973jx,Jackiw:1974cv}  provides a valuable framework for the effective potential. 
In particular, advantages of this renormalization scheme include  much simpler implementations of renormalization-group (RG) improvement properties \cite{Hill:2014mqa,Sher:1988mj,Inoue:1979nn},
the ability to address complications associated with multiple scalar field models when  couplings are large enough  that some form of resummation is necessary \cite{Gildener:1975cj,Gildener:1976ih}, 
the absence of kinetic term corrections in the effective action,  and the ability to uniquely specify the effective potential from the RG functions~\cite{Chishtie:2007vd}.
Interesting phenomenological models involving CW effective potentials include (conformal) two-Higgs doublet model \cite{Hill:2014mqa,Inoue:1979nn}, and various hidden-sector models such as the real/complex singlet model \cite{Farzinnia:2013pga,Steele:2013fka} and the $U(1)'$ model \cite{Wang:2015sxe}.
However, the CW renormalization scheme  affects the RG functions of the theory~\cite{Ford:1991hw} and hence there are scheme-transformation effects on the CW-scheme couplings that must be considered when comparing to  MS benchmark values of the couplings.

In general considerations of renormalization scheme dependence, an observable calculated to $n$-loop order in a single coupling  is known to differ between schemes and  only align as $n\to\infty$ \cite{Stevenson:1980ga}. The primary emphasis in the literature is on methods to mitigate this scheme dependence  to control theoretical uncertainties at a fixed-order of perturbation theory
 \cite{Stevenson:1981vj,Maxwell:1999dv,Brodsky:2011ta,Chishtie:2015lwk,Akrami:2020fdv}.  
Within these approaches, the combined scheme dependence in the $\beta$  function and the perturbative coefficients in the observable's loop expansion is studied.  However, a key underlying assumption in these analyses is the equivalence of the $\beta$ function coefficients up to next-to-leading order between the renormalization schemes.  As discussed below, this $\beta$ function property is not upheld in the CW renormalization scheme, motivating  our investigation of the relation between the CW and MS scheme couplings.

The CW renormalization scheme was developed for  the effective potential to study symmetry-breaking mechanisms~\cite{Coleman:1973jx,Jackiw:1974cv} and is not directly used within perturbative calculations of other observables. 
However, the constraints on couplings that emerge in CW-symmetry breaking may be implicitly referenced to the CW renormalization scheme, and hence cannot be used as input into an MS-scheme observable without scheme transformation of the coupling between MS and CW renormalization schemes.   The key purpose of this paper is to develop a bridge to convert a CW-scheme coupling (e.g., as emerges from a beyond-SM CW symmetry-breaking  mechanism) to a MS scheme coupling which can then be used as input into MS-scheme observables.

In  Section~\ref{GCW_sec} we formulate a generalization of the CW renormalization scheme that provides greater flexibility for model building and develop methods for transformation of scalar couplings between the generalized-Coleman-Weinberg (GCW) and MS schemes.  In Section~\ref{convert_sec}, numerical effects of the 
MS-GCW coupling scheme transformations are studied in detail for $\lambda\Phi^4$ theory with global $O(4)$ symmetry (the scalar sector of the SM) up to six-loop order in the MS-scheme RG functions~\cite{Kleinert:1991rg,Kompaniets:2017yct}. The availability of RG functions to this high-loop order enables a systematic study of  loop effects in the scheme transformation.   We find that the numerical effects of scheme transformation on the coupling can be significant (order of 10\%)  within the available parameter space, and hence for accurate phenomenology it is important to account for coupling scheme transformation effects in models that employ the CW renormalization scheme within their symmetry-breaking mechanisms.

\section{Generalized Coleman-Weinberg Renormalization Scheme}
\label{GCW_sec}
In its original form, the Coleman-Weinberg (CW) renormalization scheme is defined from  the following condition for the effective potential for $O(N)$ globally-symmetric  scalar field theory~\cite{Coleman:1973jx,Jackiw:1974cv}  
\begin{equation}
\left.\frac{d^4V_{eff}}{d \Phi^4}\right\vert_{\mu^2=\Phi^2}=24\lambda\,,~\Phi^2=\sum_{i=1}^N\phi_i\phi_i\,,
\label{CW_V_eff}
\end{equation}
where $\mu$ is the CW renormalization scale and condition \eqref{CW_V_eff}  is chosen to align with the definition of the tree-level Lagrangian.  However, as first noted in Ref.~\cite{Ford:1991hw}, the CW renormalization scale $\mu$ can be related to the minimal-subtraction (MS) renormalization scale $\tilde\mu$ via  
\begin{equation}
\lambda\left(\tilde\mu\right)/\tilde\mu^2=1/\mu^2\,,
\label{CW_scale}
\end{equation}
where $\lambda\left(\tilde\mu\right)$ is the MS-scheme running coupling.  
Eq.~\eqref{CW_scale} allows conversion between MS and CW renormalization schemes. 
It is evident that \eqref{CW_scale} will map typical MS-scheme effective potential logarithms $\log\left(\lambda\Phi^2/\mu^2\right)$ into CW-scheme effective potential logarithms $\log\left(\Phi^2/\mu^2\right)$.  However, \eqref{CW_scale} also implies that the RG functions in the CW and MS schemes will be different~\cite{Chishtie:2007vd,Ford:1991hw}:
\begin{equation}
\mu\frac{d\lambda}{d\mu}=\beta(\lambda)=\frac{\tilde\beta(\lambda)}{1-\frac{\tilde\beta(\lambda)}{2\lambda}}\,,
~~\tilde\mu\frac{d\lambda}{d\tilde\mu}=\tilde\beta(\lambda)\,,
\label{beta_transform}
\end{equation}
where $\tilde\beta$ denotes the MS-scheme beta function and $\beta$ denotes the CW-scheme.
A perturbative expansion of \eqref{beta_transform} provides the relation between the coefficients of the RG functions in the MS and CW schemes 
\begin{gather}
\tilde \beta(\lambda)=\sum_{k=2}^\infty \tilde b_k \lambda^k\,,
 \beta(\lambda)=\sum_{k=2}^\infty  b_k \lambda^k\,,
\end{gather}
with the first few terms given by~\cite{Chishtie:2007vd}
\begin{equation}
b_2=\tilde b_2\,,~b_3=\tilde b_3+\frac{1}{2}\tilde b_2^2\,,~b_4=\tilde b_4+\tilde b_2\tilde b_3+\frac{1}{4}\tilde b_2^3\,.
\label{beta_transform_coeffs}
\end{equation}
A notable feature of the CW scheme is the deviation from the MS-scheme $\beta$ function beginning at two-loop order (i.e., $b_3$).
Relations similar to \eqref{beta_transform} and \eqref{beta_transform_coeffs} exist for the anomalous field dimension RG function~\cite{Chishtie:2007vd}.  It can  be verified that the two-loop CW-scheme effective potential \cite{Jackiw:1974cv} satisfies the RG equation containing the CW-scheme RG functions  \cite{Chishtie:2007vd}.  

The CW-MS scheme transformation expression \eqref{CW_scale} represents a dynamical rescaling between the renormalization scale in the two schemes governed by the MS running coupling.  Given an MS coupling  $\lambda\left(\tilde\mu\right)$ at scale  $\tilde \mu$, the corresponding CW-scale $\mu$ can be calculated via \eqref{CW_scale} and the corresponding CW coupling is then given by $\lambda(\mu)$.    In principle the process can be inverted: given a CW 
coupling $\lambda\left(\mu\right)$ at scale $\mu$, \eqref{CW_scale} can be solved for $\tilde\mu$ and then the corresponding MS coupling is given by $\lambda\left(\tilde \mu\right)$.  The couplings in the two schemes match at a special scale $\mu_*$ where 
\begin{equation}
\lambda\left(\mu_*\right)=1\,,~\mu=\tilde\mu=\mu_*\,\longrightarrow \lambda(\mu)=\lambda\left(\tilde \mu\right)=1\,.
\label{CW_align}
\end{equation}
Since the MS-scheme scalar couplings increase with increasing energy scale in the perturbative regime $\lambda\left(\tilde\mu\right)<1$, Eq.~\eqref{CW_scale} implies that for $\mu<\mu_*$, 
$\mu>\tilde\mu$ and therefore the CW-scheme  coupling  is  naturally enhanced compared to the MS-scheme coupling.  It is also clear that the difference between the scales could be significant when $\lambda\left(\tilde \mu\right)\ll 1$, corresponding to a large dynamical rescaling between $\mu$ and $\tilde\mu$.

The alignment of the two schemes at the scale $\mu_*$ represented by  \eqref{CW_align} suggests a natural extension of the CW scheme \eqref{CW_scale} 
\begin{equation}
\lambda\left(\tilde\mu\right)/\tilde\mu^2=\lambda_0/\mu^2\,,
\label{GCW_scale}
\end{equation}
defining the generalized Coleman-Weinberg  (GCW) scheme transformation.    This generalization does not alter the relationship between the beta  functions in the two schemes given in \eqref{beta_transform} and  will still map MS-scheme effective potential logarithms into CW-scheme logarithms.\footnote{This freedom is also anticipated in \cite{Ford:1991hw}.}
  The parameter $\lambda_0$ then characterizes the matching between the two schemes:
\begin{equation}
\lambda\left(\mu_*\right)=\lambda_0\,,~\mu=\tilde\mu=\mu_*\,\longrightarrow \lambda(\mu)=\lambda\left(\tilde \mu\right)=\lambda_0\,,
\label{GCW_align}
\end{equation}
providing greater flexibility  for model building where it may be desirable to match the CW and MS schemes at a coupling $\lambda_0$ and a scale $\mu_*$ that emerge from a UV completion of the SM scalar sector.  
Similar to the case where $\lambda_0=1$, for scalar couplings that increase with increasing energy scale,  there will be a natural enhancement of the CW coupling compared to the MS coupling for $\mu<\mu_*$ and a natural suppression for $\mu>\mu_*$.

The matching condition which determines $\lambda_0$ should originate from a physics condition  
 while there is no clear physics behind the original CW matching condition (i.e.,~both MS coupling and CW coupling are matched at $\lambda_0=1)$.
 There are a variety of possibilities for fixing the model-dependent parameter $\lambda_0$.  
For example, Eq.~\eqref{beta_transform} implies that the MS and GCW schemes will share the same fixed points (both UV and IR) that could define $\lambda_0$.  Similarly, the anomalous dimension will have the same zeroes in the two schemes thereby providing a value for $\lambda_0$.  Because the GCW scheme transformation maps MS to the  CW  forms of the effective potentials, $\lambda_0$  can be constrained by matching the effective potentials at a particular loop order.  A UV boundary condition on $\lambda_0$ could also emerge from a UV completion (e.g., asymptotic safety \cite{Shaposhnikov:2009pv,Mann:2017wzh,Molinaro:2018kjz,Wang:2018yer,Sannino:2019sch}). 
 For example, an asymptotic safety condition that provides a coupling referenced to a UV scale $\mu_*$ then provides a matching condition to define $\lambda_0$ for a CW symmetry-breaking mechanism which results in a GCW-scheme coupling at the symmetry-breaking scale.  The connection to low-energy MS-scheme observables  (including other couplings in the model) then requires GCW-MS scheme conversion of the coupling. 
 Finally, $\lambda_0$ could be determined through  approaches for physical observables such as the principle of minimal sensitivity for a particular observable \cite{Stevenson:1981vj} or the recently developed principle of observable effective matching \cite{Chishtie:2020ac}.

 Dimensionless CW and MS  scales associated with the GCW matching condition \eqref{GCW_align} are defined by
\begin{equation}
\xi=\mu/\mu_*\,,~\tilde\xi=\tilde\mu/\mu_*\,,~\lambda(\xi=1)=\lambda(\tilde\xi=1)=\lambda_0\,,
\label{boundary_condition}
\end{equation}
providing a boundary condition for the RG equations \eqref{beta_transform} expressed in terms of the dimensionless scales
\begin{gather}
\tilde\xi\frac{d\lambda}{d\tilde\xi}=\tilde\beta(\lambda)\,,~\lambda(\tilde\xi=1)=\lambda_0
\label{MS_RG}
\\
\xi\frac{d\lambda}{d\xi}=\beta(\lambda)=\frac{\tilde\beta(\lambda)}{1-\frac{\tilde\beta(\lambda)}{2\lambda}}\,,~\lambda(\xi=1)=\lambda_0
\,.
\label{CW_RG}
\end{gather}
Similarly, the GCW scheme transformation \eqref{GCW_scale} expressed in terms of dimensionless scales is 
\begin{equation}
\lambda\left(\tilde\xi\right)/\tilde\xi^2=\lambda_0/\xi^2\,.
\label{GCW_dim_scale}
\end{equation}

Geometric aspects of the dynamical rescaling  relationship \eqref{GCW_dim_scale} between the GCW and MS schemes is a generic feature, and in principle can be interpreted in a non-perturbative context.  Given an MS-scheme  RG trajectory and a scale $\tilde\xi^2=\tilde\Lambda^2$, the dynamical rescaling  $\Lambda^2= \tilde\Lambda^2\lambda_0/\lambda\left(\tilde\Lambda^2\right)$  and the associated GCW coupling 
$\lambda_{\rm CW}
=\lambda\left(\Lambda^2\right)
$ 
is illustrated geometrically in Fig.~\ref{geometry_figure} for an RG trajectory typical of the detailed analysis presented below.  As evident from Fig.~\ref{geometry_figure}, the dynamical rescaling inevitably leads to a difference between the GCW and MS couplings, irrespective of the underlying origin (e.g., perturbative order) of the RG trajectory.  This geometrical interpretation can thus be applied to phenomenological-inspired qualitative models for the MS coupling. 

\begin{figure}[hbt]
\includegraphics[scale=0.6]{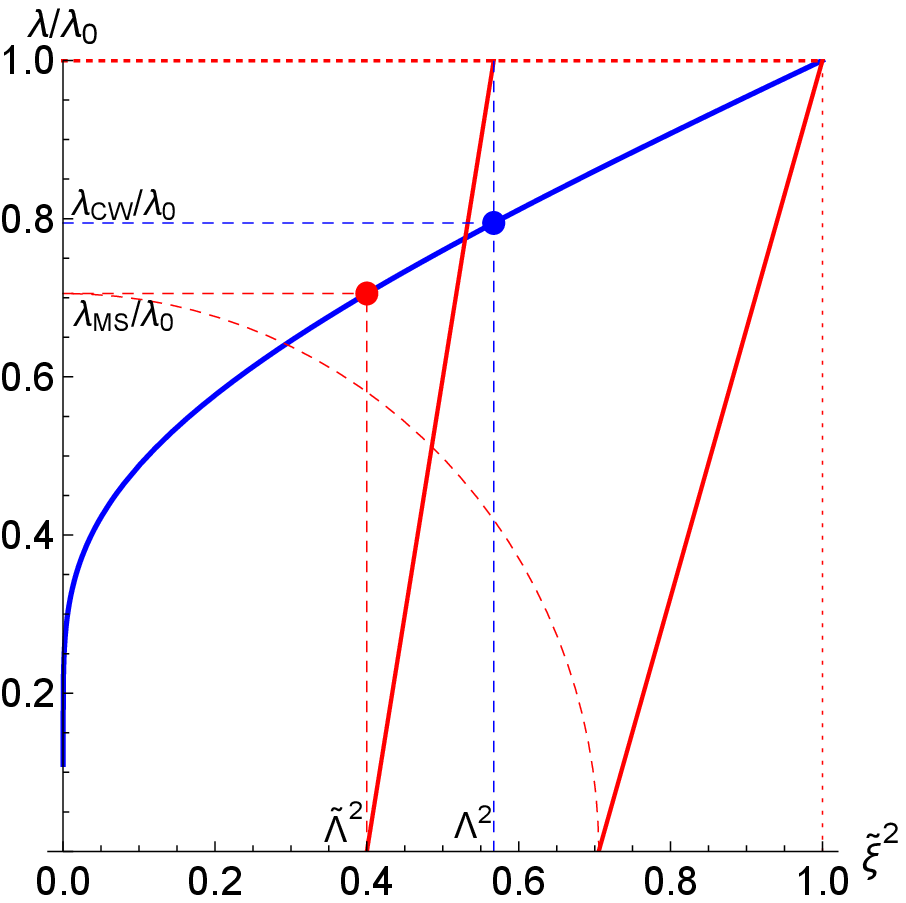}\hspace{15pt}
\includegraphics[scale=0.6]{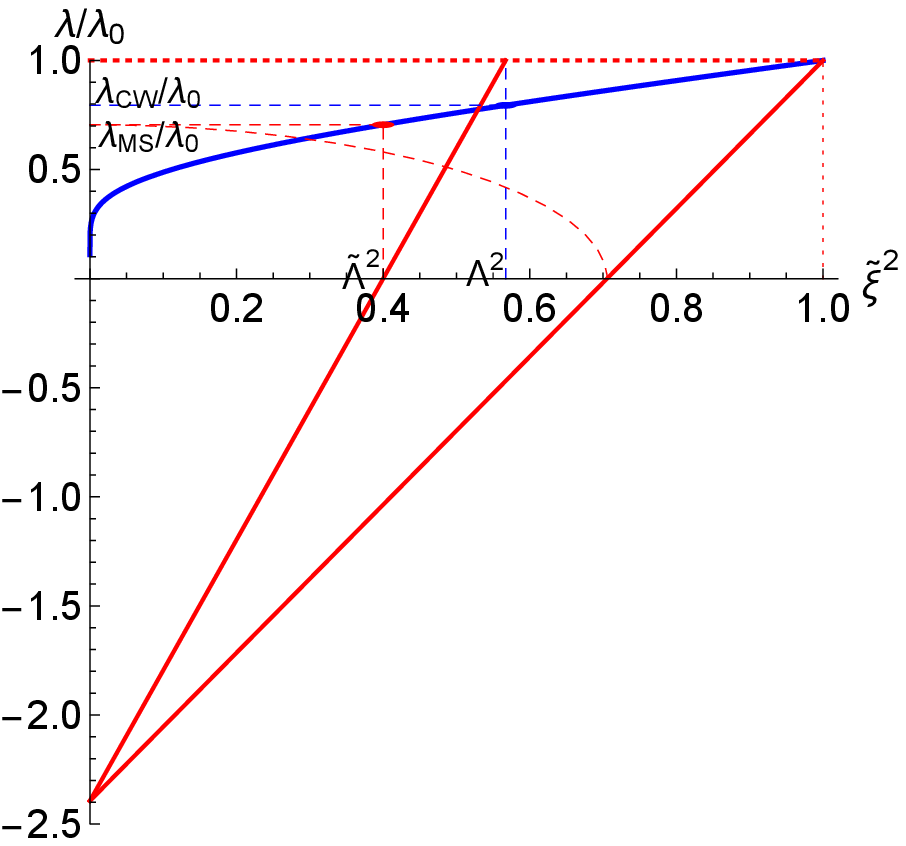}
\caption{The geometric representation of the dynamical rescaling relationship \eqref{GCW_dim_scale} between GCW scales and couplings is shown in the right panel, with a magnified region shown in the left panel.  The MS RG coupling trajectory is shown in solid blue. The MS scale $\tilde\Lambda^2$ and  associated coupling $\lambda_{\rm MS}=\lambda\left(\tilde\Lambda^2\right)$ are shown by the red circle and dashed red lines, and the $\lambda_{\rm MS}$ value of the coupling is projected to the $\tilde\xi^2$ axis by the dashed-red circle. The solid red lines, dotted red line, and coordinate axes are a geometric number-line representation of the quotient 
$\Lambda^2= \tilde\Lambda^2\lambda_0/\lambda\left(\tilde\Lambda^2\right)$, with the result identified by the intersection of the left solid-red line with the dotted red line. The dashed blue lines  and blue circle represent the GCW scale  $\Lambda^2$ and associated coupling  $\lambda_{\rm CW}=\lambda\left(\Lambda^2\right)$.
}
\label{geometry_figure}
\end{figure}


\section{Scheme Conversion of Couplings}
\label{convert_sec}

For the detailed numerical analysis between the GCW and MS schemes, the solution to the MS-scheme RG equation \eqref{MS_RG} will be denoted by $\lambda_{\MSRG}\left(\tilde\xi,\lambda_0\right)$ and the solution to the CW-scheme RG equation will be denoted by $\lambda_{\CWRG}\left(\xi,\lambda_0\right)$. 
Thus the GCW scheme transformation is founded on the MS-scheme running coupling through the dynamical scale transformation
\begin{equation}
\xi=c\left(\tilde\xi\right)=\tilde\xi \sqrt{\frac{\lambda_0}{\lambda_{\MSRG}\left(\tilde\xi,\lambda_0\right)}}\,.
\label{c_defn}
\end{equation}

For a given value of the MS scale $\tilde\xi$, the corresponding GCW-scheme coupling is obtained by calculating the corresponding CW-scale $\xi$ \eqref{c_defn}, and the resulting GCW coupling $\lambda_{\CW}$ is then given by
\begin{equation}
\lambda_{\text{CW}}\left(\tilde\xi,\lambda_0\right)= \lambda_{\MSRG}\left(c\left(\tilde\xi\right),\lambda_0\right)\,.
\label{lambda_CW}
\end{equation}
The inverse of this expression has an interesting symmetric form.  The MS coupling $\lambda_\MS$ corresponding to a CW scale $\xi$ is given by the $\tilde x$ solution of 
\begin{gather}
\lambda_\MS\left(\xi\right)=\lambda_\MSRG\left(\tilde x,\lambda_0\right)\,,
\\
\lambda_\CW\left(\tilde x\right)=\lambda_\CWRG\left(\xi,\lambda_0\right)\,.
\label{lambda_MS_con}
\end{gather}
However, \eqref{GCW_scale} and \eqref{lambda_CW} can be used to re-express \eqref{lambda_MS_con} as
\begin{equation}
\begin{split}
&\lambda_\MSRG\left(c\left(\tilde x\right),\lambda_0\right)=\lambda_\CW\left(\tilde x\right)=\lambda_\CWRG\left(\xi,\lambda_0\right)\\
&=\lambda_\MSRG\left(c^{-1}(\xi),\lambda_0\right)\rightarrow c(x)=c^{-1}(\xi)
\end{split}
\label{lambda_MS_sol}
\end{equation}
where $c^{-1}(\xi)$ is understood as the $\tilde\xi$ root of \eqref{c_defn} associated with the scale $\xi$. 
Using \eqref{lambda_MS_sol} and  \eqref{GCW_scale} leads to the final result for $\lambda_\MS$
\begin{equation}
\begin{split}
\lambda_\MS\left(\xi,\lambda_0\right)=&\lambda_\MSRG\left(\tilde x,\lambda_0\right)=\lambda_\CWRG\left(c\left(\tilde x\right),\lambda_0\right)\\
=&\lambda_\CWRG\left(c^{-1}\left(\xi\right),\lambda_0 \right)\,,
\end{split}
\label{lambda_CW_con}
\end{equation}
which has a form symmetric  to \eqref{lambda_CW}.  Thus,  Eq.~\eqref{lambda_MS_con}  performs the mapping of the MS to GCW coupling and  
 Eq.~\eqref{lambda_CW_con} performs the opposite.

In principle, the transformation of couplings between the GCW and MS schemes can be performed via a numerical solution for the MS-coupling within \eqref{lambda_CW} in cases of higher-loop beta functions where an analytic solution does not exist.  However, this purely numerical approach has some disadvantages.  First, it can be difficult to sample extreme ranges of the coupling parameter space.  For small couplings,  \eqref{GCW_dim_scale} leads to large dynamical hierarchies in $\xi$ and $\tilde \xi$, and hence a small change in $\tilde\xi$ leads to a large change in $\xi$.  For large $\lambda$ the perturbative series may begin to have poor convergence and because it is necessary to do a numerical solution from the boundary condition $\lambda_0$, the coupling could enter a
 non-perturbative regime.  Thus it is necessary to go beyond purely numerical solutions and develop other methodologies for performing  GCW-MS scheme transformation.  
 
\begin{figure}[hbt]
\includegraphics[scale=0.6]{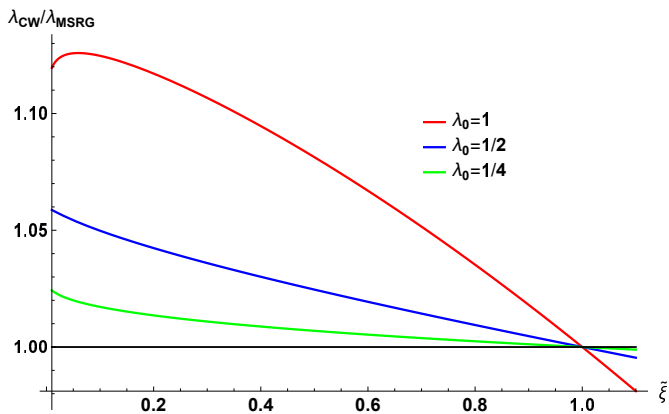}
\caption{The ratio of $\lambda_\CW$ and $\lambda_\MSRG$ is shown at one-loop order in $O(4)$ $\lambda\Phi^4$ 
as a function of the dimensionless scale $\tilde\xi$ for selected $\lambda_0$.
}
\label{one_loop_xi}
\end{figure}

The one-loop analysis for $\lambda\Phi^4$ theory with $O(4)$ global symmetry provides  a phenomenologically relevant exactly solvable case that illustrates the main features of the GCW-MS scheme transformation and motivates the methodology that will be developed for the higher-loop cases.  Using the one-loop $O(4)$ beta function \cite{Kleinert:1991rg}, the solution to \eqref{MS_RG} is
\begin{equation}
\lambda_{\MSRG}\left(\tilde\xi,\lambda_0\right)=\frac{\lambda_0}{1-\tilde b_2\lambda_0\log{\tilde\xi}}\,,~\tilde b_2=6/\pi^2\,,
\end{equation}
which combined with \eqref{lambda_CW} leads to the relationship between $\lambda_\CW$ and $\lambda_\MSRG$ shown in Figure~\ref{one_loop_xi} for different choices of $\lambda_0$.  The curves intersect at $\tilde\xi=1$ as required by  the boundary condition \eqref{boundary_condition}, and as discussed above, the GCW-MS scheme transformation leads to a natural enhancement of the CW coupling below $\tilde\xi<1$ and a suppression for $\tilde\xi>1$.  Depending on $\lambda_0$, the enhancement can be numerically significant and it is clear that a naive assumption that the CW and MS coupling are identical could introduce an error of up to order of 10\%.  Thus depending on the desired phenomenological precision of the couplings,  a careful consideration of GCW-MS scheme transformation effects may be needed. 

\begin{figure}[hbt]
\includegraphics[scale=0.6]{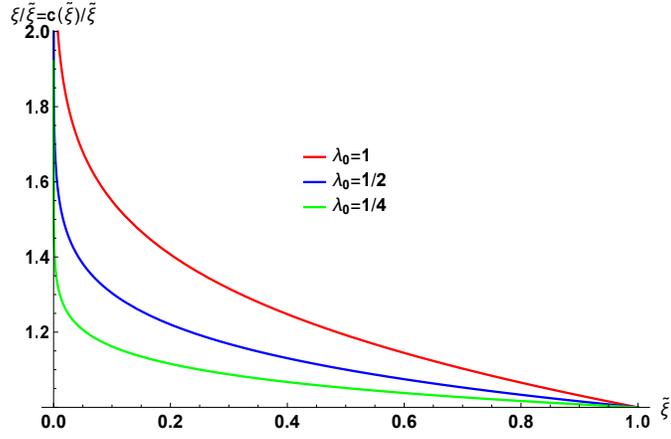}
\caption{The ratio  of $\xi/\tilde\xi=c\left(\tilde\xi\right)/\tilde\xi$ is shown at one-loop order in $O(4)$ $\lambda\Phi^4$ theory as a function of the dimensionless scale $\tilde\xi$ for selected $\lambda_0$.
  }
\label{one_loop_scale}
\end{figure}

 Figure~\ref{one_loop_scale} shows the underlying relation \eqref{c_defn} between the GCW scale $\tilde \xi$ and MS scale $\xi$ for different choices of $\lambda_0$. As discussed above,  $\xi$ is enhanced compared to $\tilde\xi$ leading to the natural enhancement of $\lambda_\CW$.  However, although $\xi>\tilde\xi$, it is apparent that $\xi<1$ for $\tilde\xi<1$ so that 
it is not necessary to evolve the MS couplings above $\lambda_0$, providing some control over perturbative convergence.  
This $\tilde\xi<\xi<1 $ property can be qualitatively understood from Fig.~\ref{geometry_figure} as geometrically arising from the convexity of the coupling RG trajectory.
The  non-linear dynamical rescaling between $\xi$ and $\tilde\xi$ in Fig.~\ref{one_loop_scale} illustrates why the GCW-MS scheme  transformation can become significant.

\begin{figure}[hbt]
\includegraphics[scale=0.6]{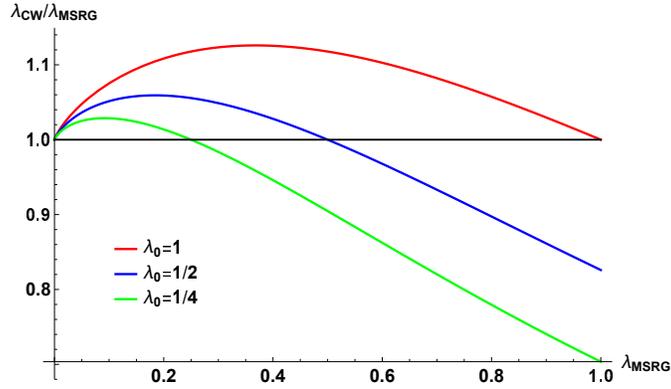}
\caption{The relationship \eqref{lambda_CW_R1} between the GCW coupling $\lambda_\CW$ and MS coupling $\lambda_\MSRG$ is shown at one-loop order in $O(4)$ $\lambda\Phi^4$ theory for selected $\lambda_0$.
 }
\label{one_loop_coupling}
\end{figure}

 Fig.~\ref{one_loop_scale} illustrates one of the challenges of the direct application of the GCW scheme transformation \eqref{lambda_CW}.  For $\tilde\xi\ll1$, a small variation in $\tilde\xi$ leads to a large change in $\xi$, thus making it difficult to sample a full range of coupling parameter space.    However, the one-loop case provides a way forward by solving the MS  RG equation \eqref{MS_RG} to relate the MS-scheme coupling at two scales
 \begin{equation}
 \lambda\left(\tilde\xi_1\right)=
 \frac{\lambda\left(\tilde\xi_2\right)}{1-\tilde b_2\lambda\left(\tilde\xi_2\right)\log{\left(\tilde\xi_1/\tilde\xi_2\right)}}\,,
\label{two_scale_one_loop}
 \end{equation} 
 where the notation has been compressed so that $\lambda\left(\tilde\xi_i\right)=\lambda_\MSRG\left(\tilde\xi_i,\lambda_0\right)$  in \eqref{two_scale_one_loop}.
By choosing $\tilde\xi_1=\xi=c(\tilde\xi)$ and $\tilde\xi_2=\tilde\xi$ related through the GCW scheme transformation \eqref{GCW_dim_scale}, Eq.~\eqref{two_scale_one_loop} provides a direct relation between $\lambda_\CW$ and $\lambda_\MSRG$
\begin{equation}
\begin{split}
\lambda_\CW&=\lambda_\MSRG ~S_0(w)\,,~S_0(w)=1/w\,,
\\
w&=1-\frac{\tilde b_2}{2}\lambda_\MSRG\,\log{\left(\frac{\lambda_0}{\lambda_\MSRG}\right)}
\end{split}
\label{lambda_CW_R1}
\end{equation}
where the functional arguments of $\lambda_\CW$ and $\lambda_\MSRG$ have been suppressed for simplicity.  Eq.~\eqref{lambda_CW_R1} provides an analytic  relation between the CW and MS couplings without requiring a solution of the MS RG equation, addressing the challenges of the direct numerical  approach  outlined above.  Furthermore, because we do not to solve \eqref{MS_RG}, $\lambda_0$ could be in a non-perturbative regime, and the   applicability of \eqref{lambda_CW_R1} is simply constrained by perturbative convergence of higher-loop contributions as outlined below.
Fig.~\ref{one_loop_coupling} shows how the relation \eqref{lambda_CW_R1}
between the GCW- and MS-scheme couplings can now be examined across a wider range of coupling parameter space (e.g., compare with Fig.~\ref{one_loop_scale}) and  the scale has been extended into a region of slow perturbative convergence to show the required $\lambda_\CW/\lambda_\MSRG=1$ intersection of each curve at $\lambda_\MSRG=\lambda_0$ from the matching condition  \eqref{boundary_condition}.  A natural enhancement of the CW coupling now occurs for $\lambda_\MSRG<\lambda_0$ and a natural suppression for  $\lambda_\MSRG>\lambda_0$.   

If the expression \eqref{two_scale_one_loop} is expanded as a series in $\lambda\left(\tilde\xi_2\right)$ then it has the form of a leading-logarithm ($LL$) summation, and hence following Ref.~\cite{Chishtie:2018ipg}  a higher-loop extension of  \eqref{two_scale_one_loop} can be found by including sub-leading logarithms in the series solution
\begin{gather}
 \lambda\left(\tilde\xi_1\right)= \lambda_2\left[
 1+T_{1,1}L\lambda_2+\lambda_2^2\left(T_{2,1}L+T_{2,2}L^2\right)+\ldots\right]
 \nonumber\\
L=\log{\left(\tilde\xi_1/\tilde\xi_2\right)}\,,~\lambda_2=\lambda\left(\tilde\xi_2\right)\,,
\label{lambda_T}
\end{gather}
which can be rearranged as sums of $N^nLL$ terms as
\begin{equation}
 \lambda\left(\tilde\xi_1\right)=\sum_{n=0}^\infty \lambda_2^{n+1}S_n\left(\lambda_2 L \right)
\,,~S_n(u)=\sum_{k=1}^\infty T_{n+k,k}u^k \,.
\label{NLL}
\end{equation}
The notation has again been compressed so that $\lambda\left(\tilde\xi\right)=\lambda_\MSRG\left(\tilde\xi,\lambda_0\right)$ and will be restored later for clarity.
The requirement that \eqref{NLL} is independent of the scale $\tilde\xi_2$ \cite{Chishtie:2018ipg}
\begin{equation}
0=\frac{d}{d\tilde\xi_2} \lambda\left(\tilde\xi_1\right)=\left(\tilde\xi_2\frac{\partial}{\partial\tilde\xi_2}+\tilde\beta\left(\lambda_2\right)\frac{\partial}{\partial\lambda_2}
\right)\lambda\left(\tilde\xi_1\right)
\end{equation}
provides an RG equation defining the $S_n$. For example, at $LL$  order the RG equation for $S_0$ is
\begin{equation}
0=(1-\tilde b_2 u)\frac{dS_0}{du} -\tilde b_2S_0\,,~S_0(0)=1
\label{S0_defn}
\end{equation}
where the boundary condition for $S_0$ ensures that \eqref{NLL} is self-consistent when $\tilde\xi_1=\tilde\xi_2$. The solution to \eqref{S0_defn} is
\begin{equation}
S_0(w)=1/w\,,~w=1-\tilde b_2u 
\end{equation}
and hence
\begin{equation}
 \lambda\left(\tilde\xi_1\right)=\lambda_2 S_0\left(\lambda_2 L\right)=\frac{\lambda_2}{1-\tilde b_2\lambda_2L}\,,
\end{equation}
identical to the one-loop result \eqref{two_scale_one_loop}. At $N^nLL$ order  the generalization of \eqref{S0_defn} to $n>0$ is
\begin{equation}
0=-\left(1-\tilde b_2u\right)\frac{dS_n}{du}+(n+1)\tilde b_2 S_n
%
+\sum_{k=0}^{n-1}\tilde b_{n+2-k}\left[(k+1)S_k+u\frac{dS_k}{du}\right]
\,, ~S_n(0)=0\,.
\label{Sn_defn}
\end{equation}
Since the MS beta function is known to six-loop order (i.e., up to $\tilde b_7$)   \cite{Kompaniets:2017yct} \eqref{Sn_defn} can be iteratively solved up to $S_5(u)$.  The next two solutions are 
\begin{gather}
S_1(w)=-\frac{\tilde b_3}{\tilde b_2}\frac{\log{w}}{w^2}\,,~w=1-\tilde b_2u \\
S_2(w)=\frac{\left(\tilde b_2\tilde b_4-\tilde b_3^2\right)(1-w)-\tilde b_3^2\log{w}+\tilde b_3^2\log^2{w} }{\tilde b_2^2w^3}\,,
\nonumber
\end{gather}
and the remaining solutions up to $S_5$ are too lengthy to be presented.  
The resulting  $N^nLL$ expression relating the MS-scheme coupling at the scales $\tilde \xi_1$ and $\tilde \xi_2$ is
\begin{equation}
\lambda^{(n)}\left(\tilde\xi_1\right)=\sum_{k=0}^{n}\lambda_2^{k+1}S_k(w)\,,~w=1-\tilde b_2
\lambda_2\log{\left(\tilde\xi_1/\tilde\xi_2\right)}\,,~\lambda_2=\lambda\left(\tilde\xi_2\right)\,.
\label{lambda_MS_resum}
\end{equation}
The accuracy of the $N^nLL$ approximations \eqref{lambda_MS_resum} can be checked by setting $\tilde\xi_2=1$, $\lambda_2=\lambda_0$, and then comparing against numerical solution of MS-scheme coupling at the same order. As expected, agreement tends to worsen as $\lambda_0$ increases because the truncated perturbative expansion will have slower convergence. The estimated  limitations on $\lambda$ are discussed below.
  
  \begin{figure}[hbt]
\includegraphics[scale=0.6]{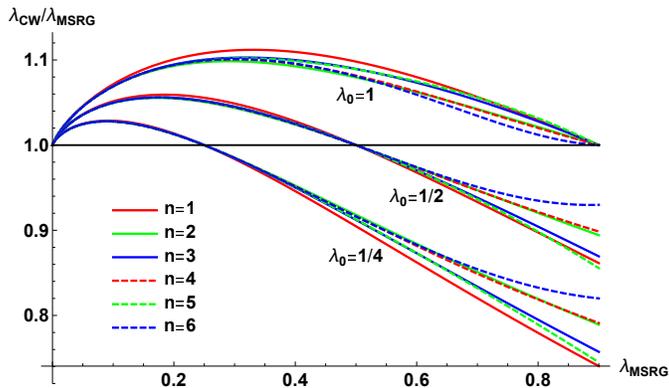}
\caption{The relationship \eqref{lambda_CW_Rn} between the GCW coupling $\lambda^{(n)}_\CW$ and MS coupling $\lambda_\MSRG$ is shown to successively higher-loop $N^nLL$  order in $O(4)$ $\lambda\Phi^4$ theory for three different $\lambda_0$. 
}
\label{multi_loop_coupling_1}
\end{figure}
  
Although the development of \eqref{lambda_MS_resum} involved an RG equation in the scale $\tilde\xi_2$, Ref.~\cite{Chishtie:2018ipg} argues that the RG equation in $\xi_1$ is contained within the $\tilde\xi_2$ RG equation for the series expansion \eqref{lambda_T}.  For the truncated series solutions \eqref{lambda_MS_resum} it can be verified that the $N^nLL$ expression self-consistently satisfies the $\tilde\xi_1$ RG equation up to $N^nLL$ order ({\it i.e.,} residual RG terms are proportional to next-order corrections).  Finally, by choosing $\tilde\xi_1=\xi$ and $\tilde\xi_2=\tilde\xi$ related through the GCW scheme transformation \eqref{GCW_dim_scale}, Eq.~\eqref{lambda_MS_resum} provides a direct $N^nLL$ relation between $\lambda_\CW$ and $\lambda_\MSRG$ 
\begin{equation}
\lambda^{(n)}_\CW=\sum_{k=0}^n\lambda^{k+1}_\MSRG \,S_k(w)\,,~w=1-\frac{\tilde b_2}{2}\lambda_\MSRG\,\log{\left(\frac{\lambda_0}{\lambda_\MSRG}\right)}\,.
\label{lambda_CW_Rn}
\end{equation}

Before showing the higher-loop scheme transformation results  it is important to note that the six-loop MS-scheme beta function contribution is surprisingly large for $O(4)$ $\lambda\Phi^4$ theory \cite{Kompaniets:2017yct} which restricts the coupling parameter space that can be explored reliably.  For example, the six-loop MS-scheme beta function has a zero at approximately $\lambda=0.93$ and hence couplings beyond this apparent fixed point should be excluded because they are in conflict with non-perturbative lattice studies of scalar field theory \cite{Luscher:1987ay,Luscher:1987ek,Luscher:1988uq,Callaway:1988ya}.  Similarly, the two- and four-loop  MS-scheme beta functions have false zeroes respectively at $\lambda=3.04$ and $\lambda=1.35$ which are safely avoided by respecting the six-loop bound.\footnote{In the normalization $\frac{\lambda}{4!}\phi^4$ instead of  $\lambda\phi^4$, the six-loop false fixed point will be rescaled from $1$ to $24\approx 8\pi$  consistent with the unitarity bound.  }

Fig.~\ref{multi_loop_coupling_1}
 compares the GCW- and MS-scheme couplings at successively higher-loop $N^nLL$ orders arising from \eqref{lambda_CW_Rn} for different choices of $\lambda_0$.  As in the exact one-loop case, a  natural enhancement of the CW coupling  occurs for $\lambda_\MSRG<\lambda_0$ and a natural suppression for  $\lambda_\MSRG>\lambda_0$.  
 As $\lambda_0$ is decreased, the enhancement for $\lambda<\lambda_0$ tends to decrease, while the suppression for $\lambda>\lambda_0$ tends to increase.  As noted before, the scheme-transformation effects can lead to significant (order of 10\%) differences in the coupling, which could be further magnified by subsequent large-distance RG running to the desired phenomenological scale.

 An interesting feature of Fig.~\ref{multi_loop_coupling_1} is the persistence of the relative enhancement (or suppression) as loop order is increased, emphasizing that the effects of the GCW-MS scheme transformation are fundamentally related to the non-linear dynamical rescaling of the renormalization scales \eqref{GCW_dim_scale} and cannot be avoided by going to higher-loop order.  A geometric representation of this dynamical rescaling and the resulting scheme transformation between couplings is shown in Fig.~\ref{geometry_figure}. 

The ideas presented in this paper can be extended to systems with multiple scalar fields and multiple couplings.  In cases where the couplings satisfy the flatness condition, the Gildener-Weinberg method   \cite{Gildener:1975cj,Gildener:1976ih} can be implemented and the single-coupling analysis of this paper can be applied without modification.   Otherwise, multi-scale RG methods  \cite{Einhorn:1983fc,Ford:1996hd,Steele:2014dsa} are required to generalize  
Eq.~\eqref{GCW_scale} to introduce an additional renormalization scale and associated parameter $\lambda_0$ for each required coupling. 



\section{Discussions: Connection between (Ultra) Strong 1st Order Phase Transition and Coleman-Weinberg Symmetry Breaking}
In this section, we  discuss the connection between a (ultra) strong first order phase transition and Coleman-Weinberg symmetry breaking.

\subsection{Coleman-Weinberg Triggering of First-Order Phase Transition}
The reason why  the Coleman-Weinberg mechanism can help to naturally realize a first order phase transition can be simply understood as follows. Imagine a model can have Coleman-Weinberg symmetry breaking. This immediately implies that at zero temperature, the system already possesses a second order phase transition. When turning on the temperature $T$, the  finite temperature contribution will always provide a positive $T^2$ contribution which will give another curvature in the effective potential at $\phi=0$. A second order phase transition at zero temperature plus a curvature at $\phi=0$ triggered by a $T^2$ term is a first order phase transition. Thus, we reach our first conclusion that any model with Coleman-Weinberg symmetry breaking at zero temperature will automatically provide a first order phase transition at finite temperature (see e.g., Ref.~\cite{Sher:1988mj}). 

\subsection{(Ultra) Strong 1st Order Phase Transition}
\label{sub_B}
Models with Coleman-Weinberg symmetry breaking at zero temperature will easily lead to a {\em strong} first order phase transition and sometimes extremely strong which implies  ultra super cooling. Cases with supercooling and ultra supercooling have been one of the frontiers of cosmology and particularly  gravitational wave research (see e.g.~\cite{Konstandin:2011dr,Ellis:2018mja,Brdar:2019qut,Ellis:2020awk}). The easy formation of a strong first order phase transition and associated supercooling in models with Coleman-Weinberg symmetry breaking was first discussed  in Ref.~\cite{Witten:1980ez}. The deep reason is that in a scale invariant theory, the tunnelling rate from the false vacuum would necessarily scale as $T^4$ times a function of the couplings. For weak couplings, the vacuum decay rate would be extremely small at arbitrarily small $T$ compared with the ordinary case that the vacuum decay rate will increase quickly when $T$ is below the critical temperature $T_c$. Thus, for the weak coupling case supercooling occurs, which also means a strong first order phase transition is very easily formed. In \cite{Witten:1980ez}, it was shown that in an ultra supercooling case, the electroweak phase transition can only happen when the universe is cooling down to the QCD scale. Eventually, the QCD chiral phase transition further triggers the electroweak phase transition. Thus, contrary to the naive expectation that weaker couplings in  models with Coleman-Weinberg symmetry breaking will lead to a weaker first order phase transition, it actually leads to a stronger first order phase transition. 

\subsection{Models with Classical Scale Invariance as a Limiting Case}
It is important to view the classical-scale invariant model with Coleman-Weinberg symmetry breaking as an important limiting case even if we study the case without classical-scale invariance. Any tree level mass term can be viewed as a term which breaks the classical scale invariance and will increase the vacuum tunnelling rate and weaken the first order phase transition. Actually, it can be shown that to obtain a strong first order phase transition, the explicit mass term cannot be too big and should only be around the Coleman-Weinberg mass predicted in the models \cite{Sher:1988mj}. 
In addition, models with classical scale invariance have their own beauty of naturalness in the sense that less engineering of the parameters in the theories are required.

The conventional Standard model without any extension is not a scale invariant theory because the explicit Higgs mass term breaks the scale invariance. Thus, the Standard Model without extensions does not apply to the discussions in Section \ref{sub_B} and there is no contradiction of the well accepted statement that the Standard Model without extensions is either a second order or weakly first order phase transition. Actually, one piece of evidence in the Standard Model also supports the above statement that models with Coleman-Weinberg symmetry breaking favors a strong first order phase transition. It is well known  that larger Higgs mass in the Standard Model (i.e., when Higgs mass is above $70\,\rm{GeV}$) will favor a second order  or weakly first order phase transition while for smaller Higgs mass will in general lead to a stronger first order phase transition \cite{Sher:1988mj}. 
The classical scale invariant theory viewed as a limiting case of small Higgs mass is consistent with this conclusion.

\subsection{Coleman-Weinberg Renormalization Scheme and Gildener-Weinberg Technique}
One of the reasons for the importance of the  Coleman-Weinberg renormalization scheme is its deep connection 
 to the Gildener-Weinberg technique \cite{Gildener:1976ih}. As stated above, models with Coleman-Weinberg symmetry breaking is very important and naturally  provides a strong first order phase transition. However,  it is well known that the Standard Model by itself cannot have Coleman-Weinberg symmetry breaking. The deep reason is even if we get rid of the explicit Higgs mass term, the vacuum will be destabilized by the large top Yukawa coupling which is unbounded from below. Thus, an extension of the scalar sector of the Standard Model is inevitable and will necessarily  lead to a model with multi-scalar fields. To study Coleman-Weinberg symmetry breaking in models with multiple scalar fields, the main technique is known as Gildener-Weinberg method \cite{Gildener:1976ih}. The advantage of this method is the multiple scalar fields can be treated as a single scalar field case which significantly simplifies the analysis. In particular, when the couplings are not too small where renormalization group (RG) analysis is required, the Gildener-Weinberg method is necessary to perform the RG analysis. However, the Gildener-Weinberg method is only rigorously consistent in the Coleman-Weinberg scheme. In conclusion, the Coleman-Weinberg scheme is extremely important when studying Coleman-Weinberg symmetry breaking in models with multiple scalar fields due to the Gildener-Weinberg method. Thus, a rigorous treatment of scheme transformation between MS and Coleman-Weinberg scheme as provided in this paper is also very meaningful.

\section{Conclusions}

In this paper, the GCW scheme  generalizing the Coleman-Weinberg renormalization  scheme \cite{Coleman:1973jx,Jackiw:1974cv,Ford:1991hw}  has been developed and GCW-MS scheme transformation of the coupling has been analyzed for  $O(4)$ globally-symmetric 
$\lambda\Phi^4$ theory up to six-loop order. 

The key messages of our paper are:
\begin{itemize}
\item Dynamical rescaling of renormalization scales leads to the transformation of couplings between generalized Coleman-Weinberg (GCW) and MS schemes.  The transformation of the GCW coupling to MS-scheme then provides a bridge to determine MS-scheme observables. The dynamical rescaling and GCW-MS scheme transformation can be conceptualized in the geometric representation of Fig.~\ref{geometry_figure}.

\item Effects of GCW-MS scheme transformation on the coupling can be large (order of 10\%) and can therefore have important phenomenological consequences if the GCW-scheme coupling is naively identified with the MS-scheme coupling within an observable (e.g., nucleation rate in first order phase transition scales as an exponential function of the scalar quartic coupling \cite{Dine:1992wr}).

\item Efficient methodologies have been developed to perform the GCW-MS scheme transformation, where the key parameter in the GCW-MS scheme conversion is $\lambda_0$ where the couplings in the two schemes align at a common energy scale $\mu_*$
\end{itemize}

\begin{acknowledgments}
The work is supported by the Natural Sciences and Engineering Research Council of Canada (NSERC) and Danish National Research Foundation under the grant DNRF:90. We are grateful to D.G.C. McKeon for valuable insights on Ref.~\cite{Chishtie:2018ipg}.
\end{acknowledgments}
 


\begin{thebibliography}{99}

\bibitem{Coleman:1973jx}
  S.~R.~Coleman and E.~J.~Weinberg,
  Phys.\ Rev.\ D {\bf 7} (1973) 1888.

\bibitem{Weinberg:1978ym}
S.~Weinberg,
Phys. Lett. B \textbf{82} (1979), 387-391.


\bibitem{Grojean:2006bp}
  C.~Grojean and G.~Servant,
  Phys.\ Rev.\ D {\bf 75} (2007) 043507


  \bibitem{Caprini:2019egz}
C.~Caprini, M.~Chala, G.~C.~Dorsch, M.~Hindmarsh, S.~J.~Huber, T.~Konstandin, J.~Kozaczuk, G.~Nardini, J.~M.~No and K.~Rummukainen, \textit{et al.}
JCAP \textbf{03} (2020)  024.

\bibitem{Sher:1988mj}
  M.~Sher,
  Phys.\ Rept.\  {\bf 179} (1989) 273.



\bibitem{Sannino:2015wka}
  F.~Sannino and J.~Virkaj\"arvi,
  Phys.\ Rev.\ D {\bf 92} (2015)  045015.

\bibitem{Huang:2020bbe}
  W.~C.~Huang, F.~Sannino and Z.~W.~Wang,
  arXiv:2004.02332 [hep-ph].

\bibitem{Hill:2014mqa}
  C.~T.~Hill,
  Phys.\ Rev.\ D {\bf 89} (2014) 073003.

\bibitem{Steele:2012av}
  T.~G.~Steele and Z.~W.~Wang,
  Phys.\ Rev.\ Lett.\  {\bf 110} (2013) 151601.

\bibitem{Meissner:2006zh}
  K.~A.~Meissner and H.~Nicolai,
  Phys.\ Lett.\ B {\bf 648} (2007) 312.

\bibitem{Elias:2003zm}
  V.~Elias, R.~B.~Mann, D.~G.~C.~McKeon and T.~G.~Steele,
  Phys.\ Rev.\ Lett.\  {\bf 91} (2003) 251601.

\bibitem{Dermisek:2013pta}
  D.~Chway, T.~H.~Jung, H.~D.~Kim and R.~Dermisek,
  Phys.\ Rev.\ Lett.\  {\bf 113} (2014)  051801.

\bibitem{Kanemura:2016tan}
  S.~Kanemura, K.~Kaneta, N.~Machida, S.~Odori and T.~Shindou,
  Phys.\ Rev.\ D {\bf 94} (2016) 015028


\bibitem{Steele:2013fka}
  T.~G.~Steele, Z.~W.~Wang, D.~Contreras and R.~B.~Mann,
  Phys.\ Rev.\ Lett.\  {\bf 112} (2014) 171602.

\bibitem{Wang:2015cda}
  Z.~W.~Wang, T.~G.~Steele, T.~Hanif and R.~B.~Mann,
  JHEP {\bf 1608} (2016) 065.


\bibitem{Agrawal:2019bpm}
  P.~Agrawal, D.~Saha, L.~X.~Xu, J.~H.~Yu and C.~P.~Yuan,
Phys. Rev. D \textbf{101} (2020)  075023 .

\bibitem{Noble:2007kk}
  A.~Noble and M.~Perelstein,
  Phys.\ Rev.\ D {\bf 78} (2008) 063518.

\bibitem{CEPC-SPPCStudyGroup:2015csa}
 M.~Ahmad {\it et al.},
``CEPC-SPPC Preliminary Conceptual Design Report. 1. Physics and Detector,''
  IHEP-CEPC-DR-2015-01,
 IHEP-TH-2015-01, IHEP-EP-2015-01.

\bibitem{Liu:2018peg}
  T.~Liu, K.~F.~Lyu, J.~Ren and H.~X.~Zhu,
  Phys.\ Rev.\ D {\bf 98} (2018) no.9,  093004



\bibitem{Jackiw:1974cv}
  R.~Jackiw,
  Phys.\ Rev.\ D {\bf 9} (1974) 1686.




\bibitem{Inoue:1979nn}
  K.~Inoue, A.~Kakuto and Y.~Nakano,
  Prog.\ Theor.\ Phys.\  {\bf 63} (1980) 234.

\bibitem{Chishtie:2007vd}
  F.~A.~Chishtie, D.~G.~C.~McKeon and T.~G.~Steele,
  Phys.\ Rev.\ D {\bf 77} (2008) 065007.


\bibitem{Farzinnia:2013pga}
  A.~Farzinnia, H.~J.~He and J.~Ren,
  Phys.\ Lett.\ B {\bf 727} (2013) 141



\bibitem{Wang:2015sxe}
  Z.~W.~Wang, F.~S.~Sage, T.~G.~Steele and R.~B.~Mann,
  J.\ Phys.\ G {\bf 45} (2018) 095002.

\bibitem{Ford:1991hw}
  C.~Ford and D.~R.~T.~Jones,
  Phys.\ Lett.\ B {\bf 274} (1992) 409,
   Erratum: [Phys.\ Lett.\ B {\bf 285} (1992) 399].

  
 \bibitem{Kleinert:1991rg}
  H.~Kleinert, J.~Neu, V.~Schulte-Frohlinde, K.~G.~Chetyrkin and S.~A.~Larin,
  Phys.\ Lett.\ B {\bf 272} (1991) 39,
   Erratum: [Phys.\ Lett.\ B {\bf 319} (1993) 545].
   
   \bibitem{Kompaniets:2017yct}
  M.~V.~Kompaniets and E.~Panzer,
  Phys.\ Rev.\ D {\bf 96} (2017)  036016.
  
  \bibitem{Shaposhnikov:2009pv}
  M.~Shaposhnikov and C.~Wetterich,
  Phys.\ Lett.\ B {\bf 683} (2010) 196



\bibitem{Mann:2017wzh}
  R.~Mann, J.~Meffe, F.~Sannino, T.~Steele, Z.~W.~Wang and C.~Zhang,
  Phys.\ Rev.\ Lett.\  {\bf 119} (2017) 261802.

\bibitem{Molinaro:2018kjz}
  E.~Molinaro, F.~Sannino and Z.~W.~Wang,
  Phys.\ Rev.\ D {\bf 98} (2018) 115007.

\bibitem{Wang:2018yer}
  Z.~W.~Wang, A.~Al Balushi, R.~Mann and H.~M.~Jiang,
  Phys.\ Rev.\ D {\bf 99} (2019) 115017.

\bibitem{Sannino:2019sch}
  F.~Sannino, J.~Smirnov and Z.~W.~Wang,
  Phys.\ Rev.\ D {\bf 100} (2019) 075009.
  
  \bibitem{Stevenson:1981vj}
  P.~M.~Stevenson,
  Phys.\ Rev.\ D {\bf 23} (1981) 2916.
  
  \bibitem{Chishtie:2018ipg}
  F.~A.~Chishtie, D.~G.~C.~Mckeon and T.~N.~Sherry,
  Mod.\ Phys.\ Lett.\ A {\bf 34} (2019)  1950047.

\bibitem{Gildener:1975cj}
E.~Gildener,
  Phys.\ Rev.\ D {\bf 13} (1976) 1025.

\bibitem{Gildener:1976ih}
  E.~Gildener and S.~Weinberg,
  Phys.\ Rev.\ D {\bf 13} (1976) 3333.

\bibitem{Einhorn:1983fc}
  M.~B.~Einhorn and D.~R.~T.~Jones,
  Nucl.\ Phys.\ B {\bf 230} (1984) 261.

\bibitem{Ford:1996hd}
  C.~Ford and C.~Wiesendanger,
  Phys.\ Rev.\ D {\bf 55} (1997) 2202.

\bibitem{Steele:2014dsa}
  T.~G.~Steele, Z.~W.~Wang and D.~G.~C.~McKeon,
  Phys.\ Rev.\ D {\bf 90} (2014) 105012.





\bibitem{Stevenson:1980ga}
  P.~M.~Stevenson,
  Annals Phys.\  {\bf 132} (1981) 383.

\bibitem{Maxwell:1999dv}
  C.~J.~Maxwell,
  Nucl.\ Phys.\ Proc.\ Suppl.\  {\bf 86} (2000) 74.

\bibitem{Brodsky:2011ta}
  S.~J.~Brodsky and X.~G.~Wu,
  Phys.\ Rev.\ D {\bf 85} (2012) 034038
   Erratum: [Phys.\ Rev.\ D {\bf 86} (2012) 079903].

\bibitem{Chishtie:2015lwk}
  F.~Chishtie, D.~G.~C.~McKeon and T.~N.~Sherry,
  Phys.\ Rev.\ D {\bf 94} (2016)  054031.
  
  \bibitem{Akrami:2020fdv}
  M.~Akrami and A.~Mirjalili,
  Phys.\ Rev.\ D {\bf 101} (2020)  034007.

\bibitem{Dine:1992wr}
  M.~Dine, R.~G.~Leigh, P.~Y.~Huet, A.~D.~Linde and D.~A.~Linde,
  Phys.\ Rev.\ D {\bf 46} (1992) 550,
 
 \bibitem{Chishtie:2020ac}
F.~A.~Chishtie, 
 arXiv:2005.11783.


\bibitem{Luscher:1987ay}
  M.~Luscher and P.~Weisz,
  Nucl.\ Phys.\ B {\bf 290} (1987) 25.
  
  \bibitem{Luscher:1987ek}
  M.~Luscher and P.~Weisz,
  Nucl.\ Phys.\ B {\bf 295} (1988) 65.
  
  \bibitem{Luscher:1988uq}
  M.~Luscher and P.~Weisz,
  Nucl.\ Phys.\ B {\bf 318} (1989) 705.

\bibitem{Callaway:1988ya}
  D.~J.~E.~Callaway,
  Phys.\ Rept.\  {\bf 167} (1988) 241.



\bibitem{Konstandin:2011dr}
T.~Konstandin and G.~Servant,
JCAP \textbf{12} (2011), 009
[arXiv:1104.4791 [hep-ph]].

\bibitem{Ellis:2018mja}
J.~Ellis, M.~Lewicki and J.~M.~No,
JCAP \textbf{04} (2019), 003
[arXiv:1809.08242 [hep-ph]].

\bibitem{Brdar:2019qut}
V.~Brdar, A.~J.~Helmboldt and M.~Lindner,
JHEP \textbf{12} (2019), 158
[arXiv:1910.13460 [hep-ph]].

\bibitem{Ellis:2020awk}
J.~Ellis, M.~Lewicki and J.~M.~No,
JCAP \textbf{07} (2020), 050
[arXiv:2003.07360 [hep-ph]].

\bibitem{Witten:1980ez}
E.~Witten,
Nucl. Phys. B \textbf{177} (1981), 477-488.
  
  \end{thebibliography}
\end{document}